\documentclass[useAMS,usenatbib]{mn2e}
\usepackage[para]{threeparttable}
\usepackage{graphicx}
\usepackage{aas_macros}
\usepackage{color}
\defcitealias{Hester2008}{1}
\defcitealias{Nugent1998}{2}
\defcitealias{SewardHarnden1982}{3}
\defcitealias{Manchesteretal1982}{4}
\defcitealias{Weisskopfetal1983}{5}
\defcitealias{Kaspietal1994}{6}
\defcitealias{Sewardetal1983}{7}
\defcitealias{Largeetal1968}{8}
\defcitealias{Dodsonetal2002}{9}
\defcitealias{Aschenbachetal1995}{10}
\defcitealias{Wolszczanetal1991}{11}
\defcitealias{DermerPowale2013}{12}
\defcitealias{Sewardetal1984}{13}
\defcitealias{Parketal2010}{14}
\defcitealias{Johnstonetal1992}{15}
\defcitealias{Vink2004}{16}
\defcitealias{Manchesteretal1985}{17}
\defcitealias{Caswelletal1992}{18}
\defcitealias{Hobbsetal2004}{19}
\defcitealias{Blazeketal2006}{20}
\defcitealias{CliftonLyne1986}{21}
\defcitealias{Yuanetal2010}{22}
\defcitealias{FinleyOegelman1994}{23}
\defcitealias{Dodsonetal2002}{24}
\defcitealias{HulseTaylor1975}{25}
\defcitealias{Kovalenko1989}{26}
\defcitealias{Deweyetal1985}{27}
\defcitealias{Yaretal2004}{28}
\defcitealias{SawadaKoyama2012}{29}
\defcitealias{Gotthelfetal2013}{30}
\defcitealias{WinklerKirshner1985}{31}
\defcitealias{Beckeretal2012}{32}
\defcitealias{Vasishtetal1997}{33}
\defcitealias{HalpernGotthelf2010}{34}
\defcitealias{Sunetal2004}{35}
\defcitealias{Anetal2013}{36}
\defcitealias{Munoetal2006}{37}
\defcitealias{Satoetal2010}{38}
\defcitealias{HalpernGotthelf2010a}{39}
\defcitealias{Dibetal2008}{40}
\defcitealias{VasishtGotthelf1997}{41}
\defcitealias{GavriilKaspi2002}{42}
\defcitealias{Saakietal2013}{43}
\defcitealias{Dibetal2009}{44}
\defcitealias{Gaensleretal2005}{45}
\defcitealias{Tiengoetal2009}{46}
\defcitealias{Parketal2012}{47}
\defcitealias{Espositoetal2009}{48}
\defcitealias{Espositoetal2009a}{49}
\defcitealias{Corbeletal1999}{50}
\defcitealias{Mereghettietal2006}{51}
\defcitealias{Tendulkaretal2012}{52}
\defcitealias{Nakagawaetal2009}{53}
\title[Statistical ages and the cooling rate of nearby XDINSs]
{Statistical ages and the cooling rate of X-ray dim isolated neutron stars}

\author[R. Gill and J.S. Heyl]{Ramandeep Gill$^{1,2}$\thanks{E-mail: rgill@cita.utoronto.ca} and Jeremy S. Heyl$^1$\thanks{E-mail: heyl@phas.ubc.ca; Canada Research Chair} \\
$^1$Department of Physics and Astronomy, University of British Columbia, 6224 Agricultural Road, \\
Vancouver, British Columbia, Canada V6T 1Z1 \\
$^2$ Canadian Institute for Theoretical Astrophysics, University of Toronto, 60 St. George Street, Toronto, Ontario, M5S 3H8}
\begin{document}

\date{Accepted ---. Received ---; in original form ---}

\pagerange{\pageref{firstpage}--\pageref{lastpage}} \pubyear{2010}

\maketitle

\label{firstpage}

\begin{abstract}
The cooling theory of neutron stars is corroborated by its comparison 
with observations of thermally emitting isolated neutron stars and 
accreting neutron stars in binary systems. An 
important ingredient for such an analysis is the age of the object, 
which, typically, is obtained from the spin-down history. This age 
is highly uncertain if the object's magnetic field varies appreciably 
over time. Other age estimators, such as supernova remnant ages and 
kinematic ages, only apply to few handful of neutron stars. We 
conduct a population synthesis study of the nearby isolated 
thermal emitters and obtain their ages statistically from the 
observed luminosity function of these objects. We argue 
that a more sensitive blind scan of the galactic disk with the 
upcoming space telescopes can help to constrain the ages to higher accuracy.
\end{abstract}


\section{Introduction}
The observed thermal states of isolated neutron stars have become the primary 
source to glean useful and interesting information about the internal structure 
of neutron stars (NSs). Reconciliation of theoretical cooling curves with 
observations of nearby isolated cooling NSs is a challenging task 
\citep[see for e.g.][for a comprehensive review]{YakovlevPethick2004,Pageetal2006}. On 
the theoretical front, the problem arises from incomplete knowledge of the composition 
and equation of state (EOS) of matter in the NS core at supernuclear densities 
($\rho > 10^{14}\rmn{ g cm}^{-3}$). Many possibilities can be realized: Depending 
on the composition the EOS can be either \textit{soft} or \textit{stiff}, where 
the stiffness characterizes the compressibility of matter, and strongly depends 
on the internal degrees of freedom of the system \cite[e.g.][]{Schaabetal1996}. For 
example, for a polytropic EOS, $P=K\rho^\Gamma$, a larger adiabatic index $\Gamma$ 
yields stiffer EOS. Such an EOS generally produces larger maximum masses and 
radii of NSs than its softer counterpart. Softer EOSs can be obtained by the 
introduction of phase transitions in the theory where the core of the NS may 
be composed of boson condensates, quark or hyperonic matter. However, the presence 
of these at nuclear densities has been ruled out from the observation of a 
$1.97\pm0.04~\rmn{M}_\odot$ NS \citep{Demorestetal2010}. The cooling behavior 
of NSs depends very sensitively on the composition of the inner core, such that 
it affects the choice of the neutrino cooling process that is dominant in the first 
$10^4 - 10^5~\rmn{yrs}$ of its evolution \citep[see for e.g.][for a detailed 
description of all the neutrino emission processes in NSs]{Yakovlevetal2001}. 

X-ray observations of isolated cooling NSs have been crucial in 
their discovery \citep[see for e.g.][]{Mereghetti2011}. However, these observations 
present its own set of challenges in determining their 
cooling rate. Here, one is interested in detecting radiation emanating from the surface of the 
NS which, in the case of pulsars, is complicated by the non-thermal emission from the 
magnetosphere. Also, the non-uniform 
heating of the crust due to energetic particles accelerated in the magnetosphere render accurate 
determination of effective temperatures hard \citep[see for e.g. the review by][on 
surface emission from NSs]{Ozel2013}. The unknown composition of NS atmospheres leads to 
the overestimation of their temperatures when fitting their spectra with a blackbody 
\citep[e.g.][]{Lloydetal2003}. Apart from the
uncertainties involved in the spectral modelling of NS surface emission, uncertainties
in their distances can also contribute to poor effective temperature and luminosity
estimates. Furthermore, to place observed isolated sources on cooling curves, accurate ages 
are needed that may be hard to obtain as we discuss below.

In this study, we show that, for a large sample, ages of isolated thermal emitters can 
be derived from statistical arguments. As the sample size grows, the statistical error 
diminishes. We look at a small group of thermally emitting NSs discovered in the 
ROSAT all-sky survey in Sec. \ref{sec:M7} and derive their ages statistically in Sec. 
\ref{sec:ages}. Lastly, we discuss the possibility of improving the meager sample of 
isolated thermal emitters by detecting more such objects in the upcoming eROSITA 
all-sky survey.
\section{The Magnificent Seven}\label{sec:M7}
Nearby isolated cooling neutron stars are particularly important for confronting 
cooling models with observations. They were discovered by Einstein, ROSAT, 
and ASCA space telescopes \citep{BeckerPavlov2002}, and were further observed 
with the extremely sensitive and high resolution high-energy space telescopes Chandra 
and XMM-Newton. Among all the discovered objects, the most interesting are the seven 
radio-quiet thermally emitting isolated NSs (a.k.a. the magnificent seven, M7) 
discovered in the ROSAT all-sky survey (RASS) \citep[see for e.g.][for a review]{Haberl2007}. 
These radio-quiet objects radiate predominantly in X-rays with 
high X-ray to optical flux ratios, $f_X/f_{opt} > 10^{4 - 5}$. Their soft X-ray 
spectra are reasonably well fit by an absorbed blackbody-like spectrum with 
$kT \la 100$ eV and a hydrogen column density $n_H \sim 10^{20}\mbox{ cm}^{-2}$, 
indicating small distances $d \sim \mbox{ few }\times 100$ pc. Astrometric measurements 
of some of the member objects independently confirm the distances inferred from 
column densities (see references in Table \ref{tbl:m7table}). That the thermal 
emission is coming from majority of the stellar surface is confirmed by the small 
pulse fractions $\la 20 \%$ of the X-ray light curves. Spin periods ranging 
from 3 - 12 s have been measured for all but one (RX J1605.3+3249) 
of the M7 objects \citep[see for e.g.][Table \ref{tbl:m7table}]{Mereghetti2011}. 
This in conjunction with the measured spin-down rates, 
$\dot{P} \sim 10^{-14} - 10^{-13}\mbox{ s s}^{-1}$, yields an estimate of the polar 
magnetic field strengths $B_p\sim 10^{13}$ G and the characteristic spin-down ages 
$\tau_c\sim 10^6$ years \citep{KaplanKerkwijk2005b,KaplanKerkwijk2005a,KaplanKerkwijk2009b,
KaplanKerkwijk2009a,KaplanKerkwijk2011,KerkwijkKaplan2008}. 
Measurements of the high proper motions of three of the M7 
objects and their association thus established to the Sco OB2 complex comprising the 
Gould Belt yield kinematic and/or dynamical ages that are smaller than ages inferred from 
spin down \citep{Kaplanetal2002,Kaplanetal2007,WalterLattimer2002,Motchetal2005,
Motchetal2009,Tetzlaffetal2010,Tetzlaffetal2011,Mignanietal2013}.
 
The discrepancy between characteristic and kinematic ages strongly suggests 
that the spin-down ages are overestimates and the M7 objects in reality are much younger 
\citep[see for e.g.][]{KaplanKerkwijk2005b,KaplanKerkwijk2009b}. 
Even when considering simple cooling models \citep{HeylHernquist1998, Ponsetal2009}, 
one finds the spin-down ages to be $3-4$ times in excess of the cooling ages of $\sim 0.5~\rmn{Myr}$ 
\citep{Kaplanetal2002b}. 
\subsection{Spin-Down Ages: Poor Age Estimators}
According to the standard magnetic dipole model of pulsars \citep{ShapiroTeukolsky1983}, 
a rotating NS with a polar magnetic field spins down over time by emitting magnetic dipole 
radiation. From the rate of change of the angular frequency $\dot{\Omega}$, the spin-down age of the NS can 
be readily determined, with the assumption that the initial angular frequency is much larger than the 
present value ($\Omega_0 \gg \Omega(t)$),
\begin{equation}
\tau = \frac{\Omega}{2\|\dot{\Omega}\|}
\label{eq:sdage}
\end{equation}
This assumption is invalid in the case of CCOs as these objects are believed to have their initial periods 
very close to the current values \citep[e.g.][]{HalpernGotthelf2010}.
The spin-down law implicitly assumes that none 
of the other physical characteristics of the pulsar vary over time. This may not be the 
case and, in general, the spin-down law \citep{lynesmith2006} written as the following 
can be allowed to include variation of $B_p$, the moment of inertia $I$, and the angle 
between the rotation axis and the magnetic dipole axis $\alpha$, so that
\begin{equation}
\frac{d\Omega}{dt} = -\kappa(t)\Omega(t)^n
\end{equation}
where $\kappa(t)$ is usually assumed to be a constant and $n = 3$ is the \textit{braking index} 
for magnetic dipole braking. Any change in $\kappa$ with 
time naturally yields ages of pulsars that are in conflict with their spin-down ages; 
Generally, the spin-down age should only be taken as a rough estimate to aid in calculations.

\begin{figure*}
\centering
\includegraphics[width=0.8\textwidth]{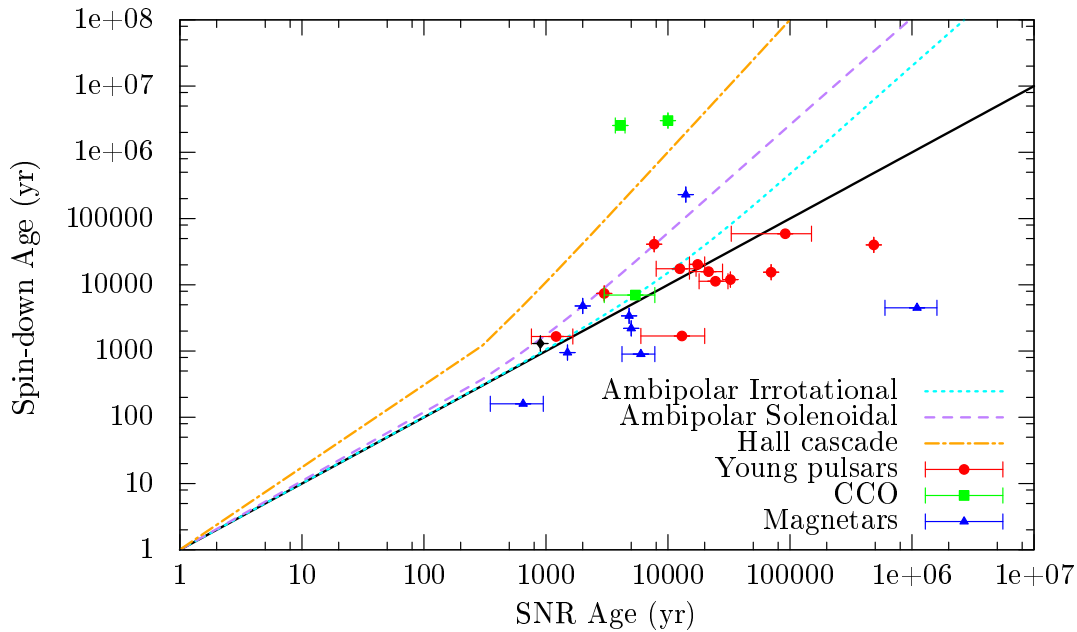}
\caption{Change in the spin-down or characteristic age of an isolated NS due to various 
magnetic field decay mechanisms, namely the ambipolar (I)rrotational 
($a = 0.01$, $\alpha = 5/4$, dotted) and 
(S)olenoidal ($a = 0.15$, $\alpha = 5/4$, dashed) modes, 
and the Hall cascade ($a = 10$, $\alpha = 1$, dot-dashed) 
\citep[See Eq. 2 - 4 of][]{Colpietal2000}. 
The solid black line denotes $t = \tau$, meaning no field decay. This assumes 
an initial field strength $B_0 = 10^{16}$ G and period $P_0 = 1$ ms. The Crab pulsar 
is shown here with a black diamond.}
\label{fig:snrVtc_fdecay}
\end{figure*}

An independent age estimate is provided by the age of the associated supernova remnant 
(SNR) or massive star cluster for younger objects. Establishing such an association for 
older NSs may prove to be difficult since SNRs fade away in $\sim 60$ kyr, and in the 
same time, due to natal kicks ($\sim 500\mbox{ km s}^{-1}$), NSs may move significantly 
far away from their birth sites \citep{Frailetal1994}. We plot the spin-down 
and the estimated SNR ages for young pulsars ($\tau < 10^5$ yrs), central compact 
objects (CCOs), and magnetars (SGRs and AXPs) along with their timing properties 
\citep[see for e.g.][for a review]{Becker2009} in Fig. \ref{fig:snrVtc_fdecay}, and it is clear 
that for many NSs, that are not the typical spin-down powered radio pulsars, 
the characteristic age is a poor age estimator. The objects that have SNR 
ages smaller than their spin-down ages can be explained by having a braking index less 
than the \textit{canonical} value, $n < 3$. An excellent example supporting this notion 
is the Vela pulsar which has a very small breaking index $n = 1.4\pm0.2$ estimated from 
an impressive 25-year long observation \citep{Lyneetal1996}, albeit under the assumption 
that $\kappa$ is still a constant. This yields a spin-down age of 25.6 kyr, making it 
appear more than twice as old as its age inferred from the standard magnetic braking 
scenario. This result is well supported by the estimated age of the Vela SNR 
($t_{\rmn{SNR}}\sim18 - 31$ kyr) \citep{Aschenbachetal1995}.

On the other hand, for objects that have spin-down ages larger than that of their true ages, 
that may be inferred from their associated SNR ages, it 
can be argued that the magnetic moments decrease in strength over time 
\citep{HeylKulkarni1998}. There are 
three main mechanisms by which magnetic fields can decay in isolated NS, namely 
Ohmic dissipation, ambipolar diffusion, and Hall drift \citep{GoldreichReisenegger1992}. 
The timescale over which the field decays substantially due to these processes 
\citep[see for e.g.][]{HeylKulkarni1998}
determines the dominating process at different stages in the evolution of an isolated NS.

An important consequence of field decay is that it leads to an overestimation of the 
real age of the NS. Following the analysis of \citep[][see Eq. 2, 3, and 4]{Colpietal2000}, we 
plot in Fig. \ref{fig:snrVtc_fdecay} the change in the spin-down age of the object over time due to the decaying strength of 
the magnetic field by the aforementioned processes.
\begin{equation}
\tau(t) = \frac{P(t)^2}{2bB(t)^2}
\end{equation}
where $b \approx 10^{-39}$ in cgs units. The effect of field decay at late times 
is apparent from the divergence of the characteristic age from the real age.
\begin{table*}
\caption{Properties of isolated NSs with SNR or massive star cluster associations}
\begin{minipage}{\textwidth}
\centering
\begin{threeparttable}
\begin{tabular}{ccccccc}
\hline
\hline
Object & $P$ & $\dot{P}$ & $t_{\rmn{sd}}$ & $t_{\rmn{snr}}$ & SNR/Cluster & References\\
 & (s) & $(10^{-15}~\rmn{s~s}^{-1})$ & $(\rmn{kyr})$ & $(\rmn{kyr})$ & & \\
\hline
\multicolumn{6}{c}{Young Pulsars$^{\dagger}$} \\
\hline
0531+21 & 0.033 &  421 & 1.3 & 0.90 & Crab Nebula & \citetalias{Hester2008}, \citetalias{Nugent1998} \\
1509-58 & 0.150 & 1536 & 1.691 & $6-20$ & MSH 15-52 & \citetalias{SewardHarnden1982}, 
\citetalias{Manchesteretal1982}, \citetalias{Weisskopfetal1983}, \citetalias{Kaspietal1994}, \citetalias{Sewardetal1983} \\
0833-45 & 0.089 & 125 & 11.3 & $18 - 31$ & Vela XYZ & \citetalias{Largeetal1968}, \citetalias{Dodsonetal2002}, \citetalias{Aschenbachetal1995} \\
1853+01 & 0.267 & 208 & 20.3 & $15-20$ & W44 & \citetalias{Wolszczanetal1991},  \citetalias{DermerPowale2013}\\
0540-69 & 0.050 & 479 & 1.66 & $0.76-1.66$ & SNR 0540-693 & \citetalias{Sewardetal1984}, \citetalias{Parketal2010} \\
1610-50 & 0.232 & 495 & 7.4 & $\sim 3.0$ & Kes 32 & \citetalias{Johnstonetal1992}, \citetalias{Vink2004} \\
1338-62 & 0.193 & 253 & 12 & $\sim 32.5$ & G308.8-0.1 & \citetalias{Manchesteretal1985},  \citetalias{Caswelletal1992} \\
1757-24 & 0.125 & 128 & 15.5 & $\ga 70$ & G5.4-1.2 & \citetalias{Manchesteretal1985}, \citetalias{Hobbsetal2004}, \citetalias{Blazeketal2006} \\
1800-21 & 0.134 & 134 & 15.8 & $15 - 28$ & W30 & \citetalias{CliftonLyne1986}, \citetalias{Yuanetal2010}, \citetalias{FinleyOegelman1994} \\
1706-44 & 0.102 & 93 & 17.5 & $8-17$ & G343.1-2.3 & \citetalias{Johnstonetal1992}, \citetalias{Dodsonetal2002} \\
1930+22 & 0.144 & 57.6 & 40 & $\sim 486$ & G57.1+1.7 &  \citetalias{HulseTaylor1975}, \citetalias{Hobbsetal2004}, \citetalias{Kovalenko1989} \\
2334+61 & 0.495 & 193 & 41 & $7.7$ & G114.3+0.3 & \citetalias{Deweyetal1985}, \citetalias{Yuanetal2010}, \citetalias{Yaretal2004} \\
1758-23 & 0.416 & 113 & 59 & $33-150$ & W28 & \citetalias{Manchesteretal1985}, \citetalias{Yuanetal2010}, \citetalias{SawadaKoyama2012} \\
\hline
\multicolumn{6}{c}{CCOs} \\
\hline
RX J0822.0-4300 & 0.122 & $0.00928\pm0.00036$ & $2.54\times10^3$ & $3.7-4.45$ & Pupppis A & \citetalias{Gotthelfetal2013}, \citetalias{WinklerKirshner1985}, 
\citetalias{Beckeretal2012} \\
1E 1207.4-5209 & 0.424 & $0.02224\pm0.00016$ & $3.02\times10^3$ & $\sim 10$ & G296.5+10.0 & \citetalias{Gotthelfetal2013}, \citetalias{Vasishtetal1997} \\
CXOU J185238.6 & 0.105 & $0.00868\pm0.00009$ & 7 & $3-7.8$ & Kes 79 & \citetalias{HalpernGotthelf2010}, \citetalias{Sunetal2004} \\
\hline
\multicolumn{6}{c}{Magnetars$^{\dagger\dagger}$} \\
\hline
CXOU J164710.2 & 10.6 & $<400$ & $>420$ & $(4\pm1)\times10^3$ & Westerlund 1${}^\star$ & \citetalias{Anetal2013}, \citetalias{Munoetal2006} \\
CXOU J171405.7 & 3.83 & $6.4\times10^4$ & 0.95 & $\sim1.5$ & CTB 37B & \citetalias{Satoetal2010}, \citetalias{HalpernGotthelf2010a} \\
1E 1841-045 & 11.78 & $3.93\times10^4$ & 4.8 & $\sim 2$ & Kes 73 & \citetalias{Dibetal2008}, \citetalias{VasishtGotthelf1997} \\
1E 2259+586 & 6.98 & 484 & 230 & 14 & CTB 109 & \citetalias{GavriilKaspi2002}, \citetalias{Saakietal2013} \\
1E 1048.1-5937 & 6.46 & $2.25\times10^4$ & 4.5 & $(1.1\pm0.5)\times10^3$ & GSH 288.3-0.5-2.8 & \citetalias{Dibetal2009}, \citetalias{Gaensleretal2005} \\
SGR 0526-66 & 8.05 & $3.8\times10^4$ & 3.4 & $\sim 4.8$ & N49 & \citetalias{Tiengoetal2009}, \citetalias{Parketal2012} \\
SGR 1627-41 & 2.59 & $1.9\times10^4$ & 2.2 & $\sim 5$ & G337.0-0.1 & \citetalias{Espositoetal2009}, \citetalias{Espositoetal2009a}, \citetalias{Corbeletal1999} \\
SGR 1900+14 & 5.2 & $9.2\times10^4$ & 0.90 & $> 6\pm1.8$ & Massive star cluster & \citetalias{Mereghettietal2006}, \citetalias{Tendulkaretal2012} \\
SGR 1806-20 & 7.6 & $7.5\times10^5$ & 0.16 & $> 0.65\pm0.3$ & Massive star cluster & \citetalias{Nakagawaetal2009}, \citetalias{Tendulkaretal2012} \\ 
\hline
\end{tabular}
\begin{tablenotes}
\setcounter{footnote}{0}
\item $^\dagger$ http://www.atnf.csiro.au/research/pulsar/psrcat (\citet{Manchesteretal2005}); 
$^{\dagger\dagger}$ http://www.physics.mcgill.ca/~pulsar/magnetar/main.html; 
$^\star$ Westerlund 1 is a massive star cluster; 
\citepalias{Hester2008} \citet{Hester2008} and references therein; 
\citepalias{Nugent1998} \citet{Nugent1998}; 
\citepalias{SewardHarnden1982} \citet{SewardHarnden1982};
\citepalias{Manchesteretal1982} \citet{Manchesteretal1982}; 
\citepalias{Weisskopfetal1983} \citet{Weisskopfetal1983};
\citepalias{Kaspietal1994} \citet{Kaspietal1994};
\citepalias{Sewardetal1983} \citet{Sewardetal1983};
\citepalias{Largeetal1968} \citet{Largeetal1968}; 
\citepalias{Dodsonetal2002} \citet{Dodsonetal2002};
\citepalias{Aschenbachetal1995} \citet{Aschenbachetal1995}; 
\citepalias{Wolszczanetal1991} \citet{Wolszczanetal1991}; 
\citepalias{DermerPowale2013} \citet{DermerPowale2013}; 
\citepalias{Sewardetal1984} \citet{Sewardetal1984}; 
\citepalias{Parketal2010} \citet{Parketal2010}; 
\citepalias{Johnstonetal1992} \citet{Johnstonetal1992}; 
\citepalias{Vink2004} \citet{Vink2004}; 
\citepalias{Manchesteretal1985} \citet{Manchesteretal1985}; 
\citepalias{Caswelletal1992} \citet{Caswelletal1992}; 
\citepalias{Hobbsetal2004} \citet{Hobbsetal2004}; 
\citepalias{Blazeketal2006} \citet{Blazeketal2006}; 
\citepalias{CliftonLyne1986} \citet{CliftonLyne1986}; 
\citepalias{Yuanetal2010} \citet{Yuanetal2010}; 
\citepalias{FinleyOegelman1994} \citet{FinleyOegelman1994}; 
\citepalias{Dodsonetal2002} \citet{Dodsonetal2002}; 
\citepalias{HulseTaylor1975} \citet{HulseTaylor1975}; 
\citepalias{Kovalenko1989} \citet{Kovalenko1989}; 
\citepalias{Deweyetal1985} \citet{Deweyetal1985}; 
\citepalias{Yaretal2004} \citet{Yaretal2004}; 
\citepalias{SawadaKoyama2012} \citet{SawadaKoyama2012}; 
\citepalias{Gotthelfetal2013} \citet{Gotthelfetal2013}; 
\citepalias{WinklerKirshner1985} \citet{WinklerKirshner1985}; 
\citepalias{Beckeretal2012} \citet{Beckeretal2012}; 
\citepalias{Vasishtetal1997} \citet{Vasishtetal1997}; 
\citepalias{HalpernGotthelf2010} \citet{HalpernGotthelf2010};  
\citepalias{Sunetal2004} \citet{Sunetal2004}; 
\citepalias{Anetal2013} \citet{Anetal2013}; 
\citepalias{Munoetal2006} \citet{Munoetal2006}; 
\citepalias{Satoetal2010} \citet{Satoetal2010}; 
\citepalias{HalpernGotthelf2010a} \citet{HalpernGotthelf2010a}; 
\citepalias{Dibetal2008} \citet{Dibetal2008}; 
\citepalias{VasishtGotthelf1997} \citet{VasishtGotthelf1997}; 
\citepalias{GavriilKaspi2002} \citet{GavriilKaspi2002}; 
\citepalias{Saakietal2013} \citet{Saakietal2013}; 
\citepalias{Dibetal2009} \citet{Dibetal2009}; 
\citepalias{Gaensleretal2005} \citet{Gaensleretal2005}; 
\citepalias{Tiengoetal2009} \citet{Tiengoetal2009}; 
\citepalias{Parketal2012} \citet{Parketal2012}; 
\citepalias{Espositoetal2009} \citet{Espositoetal2009}; 
\citepalias{Espositoetal2009a} \citet{Espositoetal2009a}; 
\citepalias{Corbeletal1999} \citet{Corbeletal1999}; 
\citepalias{Mereghettietal2006} \citet{Mereghettietal2006}; 
\citepalias{Tendulkaretal2012} \citet{Tendulkaretal2012}; 
\citepalias{Nakagawaetal2009} \citet{Nakagawaetal2009}; 
\end{tablenotes}
\end{threeparttable}
\end{minipage}
\label{tab:youngpulsars}
\end{table*}
\section{True Age Estimates of Isolated Neutron Stars}\label{sec:ages}
The M7 objects don't have any SNR or massive star cluster associations. Therefore, 
ages for these objects have been derived from their $P$ and $\dot{P}$ measurements. 
In addition, since they are nearby objects ($d \la 500$ pc), and due to their large 
proper motions, kinematic ages became a possibility and have been estimated for only 
four of the group members (see Table \ref{tbl:m7table}). In this
case, one finds that the spin-down ages are larger by a factor of $3 -
10$ than the kinematic ages. Accurate age estimates are extremely important in determining 
the cooling behavior of isolated NSs. Overestimated ages used to fit model 
cooling curves can obscure the determination of the true thermal state of these objects.
\subsection{Age Estimates from Population Synthesis}
In the following, we estimate the true ages of the M7 members by a method that is 
motivated by another method devised by \citet{Schmidt1968} and then applied by 
\citet{HuchraSargent1973} to calculate the luminosity function of field galaxies. 
The original idea is implemented as follows. An apparent magnitude limited sample 
is first obtained and it is assumed that the objects in the sample, field galaxies 
for instance, are distributed uniformly in Euclidean space, such that the 
luminosity function is independent of the distance. Then, 
to each object an \textit{accessible volume} $V_{\rmn{max}}(M)$ is assigned \citep{AvniBahcall1980}. 
This volume depends on the absolute magnitude of the object and gives a measure of the volume 
surveyed for a given object with an absolute magnitude $M$. Basically, $V_{\rmn{max}}(M)$ is 
the maximum volume in which the object would be detected had its apparent magnitude 
been equal to the limiting magnitude of the survey. The whole sample is then divided 
into bins of size $dM$ with objects having absolute magnitude in the range $[M-dM/2,M+dM/2]$. 
The luminosity function for this bin is estimated by adding the inverse of the 
accessible volumes for each object in that bin
\begin{equation}
\Phi(M) (\mbox{ Mpc}^{-3}\mbox{ mag}^{-1}) = \sum_i \frac{1}{V_{i,\rmn{max}}(M)}.
\end{equation}
This method provides a non-parametric way of estimating the luminosity function 
and it exactly reproduces the true luminosity function within statistical 
errors \citep{HartwickSchade1990}. Furthermore, this is a very general method 
which relies on only one underlying assumption that the objects are distributed 
according to their intrinsic brightnesses.

In the case of neutron stars (unlike galaxies) it is reasonable to
assume that the birthrate has been constant over the past few million
years, and furthermore, the neutron stars progress from bright to
faint luminosities as they age in the same (albeit unknown) way.  Under
these assumptions we can deduce the age of a neutron star of a given 
absolute magnitude
\begin{equation}
t(M) = \frac{1}{\beta}\int_{-\infty}^{M} \Phi(M') dM'
\end{equation}
where $\beta$ is the neutron-star birthrate per volume.
\subsection{RASS and Population Synthesis}
In the following, we develop a slight variant of the
\citet{Schmidt1968} estimator to calculate the true ages of the
members of the M7 family. The method we develop cannot be completely
model independent as the distribution of NSs in space, unlike that of galaxies
over large scales, is not uniform. Since the progenitors of NSs mainly
reside in the arms of a spiral galaxy, and for a natal kick velocity
of, say $\sim 500 \mbox{ km s}^{-1}$, the NSs only travel a distance
of $\sim 50$~pc from their birth sites within $\sim 10^5$ years. This
is small compared to the scale height of the thin disk $\sim 300$~pc
\citep{BinneyMerrifield1998}. As a result, the \citet{Schmidt1968}
estimator cannot be used here. Instead, we look at the NS progenitor
population and calculate the weight, that is used to determine the 
statistical age (see Eq. \ref{eq:age}),  for each M7 member by
counting the number of massive OB stars that are found in the
accessible volume $V_{\mathrm{max}}$ for that object.  The population
synthesis method is given in our earlier study \citep{GillHeyl2007},
and essentially requires the assumption of the luminosity function and
spatial distribution of massive OB stars in the galaxy
\citep{BahcallSoneira1980}, and the distribution of HI, which we model
as a smooth exponential disk both radially and vertically
\citep{FosterRoutledge2003}. As pointed out in \citet{Posseltetal2007}, 
local clumpiness of the ISM will affect the level of absorption. Thus, 
distances derived from assuming a homogeneous model will also be affected. This presents 
a very small degree of uncertainty in $V_\rmn{max}$, that is insignificant 
compared to other uncertainties in the model, e.g. the uncertainty in 
the SN rate used to derive statistical ages (see below).

All M7 objects were discovered by ROSAT which scanned the whole sky
with a limiting count rate of $0.015 \mbox{ cts s}^{-1}$ in the energy
range $\sim 0.12 - 2.4$ keV \citep[see][for more details]{Hunschetal1999}. The complete survey covers $92\%$ of the sky for a count
rate of $0.1 \mbox{ cts s}^{-1}$ \citep{Vogesetal1999} and has yielded
the most complete and sensitive survey of the soft X-ray sky. Therefore, it
provides a perfect flux-limited sample for our study. Next, we
calculate the weights for each object in the sample by simulating the
RASS and finding the total number of massive OB stars in the volume
$V_{\rmn{max}}$, such that the statistically estimated age is given in terms of the
typical age of their progenitors $t_{\mathrm{OB}}$,
\begin{equation}
t_{i,\rmn{stat}} \sim t_{\mathrm{OB}}\sum_{j=1}^{j=i}\frac{N_j}{N_{j\mathrm{,OB}}} 
\rightarrow t_{\mathrm{OB}}\sum_{j=1}^{j=i}\frac{1}{N_{j\mathrm{,OB}}}
\label{eq:age}
\end{equation}
where $N_j$ is the number of M7 objects in a small absolute magnitude bin of
size $dM$ centered at $M_j$. Since the sample is of marginal size,
$N_j = 1$ in this case.  Then, $1/N_{j\rmn{,OB}}$ gives the number of
massive OB stars per $j^{\rmn{th}}$ object in the sample. We first 
calculate the number of OB stars that lie in the volume $V_{\rmn{max}}$ 
for each M7 member. Then, we rank the M7 objects with respect to their 
effective temperatures with the hottest member ranked first $(i=1)$ and the 
coolest ranked last $(i=7)$. According to Eq. \ref{eq:age}, the weight for 
the first object is $N^{-1}_{1,\rmn{OB}}$, the second is 
$(N^{-1}_{1,\rmn{OB}}+N^{-1}_{2,\rmn{OB}})$, and so on.

The ages of NS progenitors are highly uncertain and are usually
obtained by estimating the main sequence turn-off ages of the massive
star cluster to which the NS may be associated \citep[see for
e.g.][for a review]{Smartt2009}. Typical ages of $\sim 3 - 15$
Myr have been estimated for the progenitors of NSs and
magnetars. \citet{Figeretal2005} report the age of the cluster of
massive stars, containing three Wolf-Rayet stars and a post
main-sequence OB supergiant, associated to the magnetar SGR $1806 -
20$ to be roughly $3.0 - 4.5$ Myr. Also, \citet{Munoetal2006} report
an age of $4\pm1$ Myr for the cluster Westerlund 1 which seems to be
the birth site of another magnetar CXOU J$164710.2 - 455216$. In yet
another study, \citet{Daviesetal2009} find the age of the cluster
associated to the magnetar SGR $1900 + 14$ to be $14\pm1$ Myr. Since
the spin-down ages of magnetars are much smaller ($\sim 10^{3 - 4}$ yr)
than that of the clusters, the notion that the cluster age reflects
the age of the progenitor, under the assumption of coevality of its
members, is a valid one. In the case of SGR $1806 - 20$ and CXOU
J$164710.2 - 455216$, both groups find that the progenitor must be a
massive star with $M > 40 M_\odot$, except in the last study where the
progenitor of SGR $1900 + 14$ is claimed to be a lower mass star with
initial MS mass of $17\pm2 M_\odot$. Notwithstanding this last result,
it has been claimed that magnetars may be the progeny of only
sufficiently massive stars ($M \ga 25 M_\odot$)
\citep{Gaensleretal2005} that would, otherwise, have resulted in the
formation of a black hole.  Although the members of the M7 family are
endowed with fields an order of magnitude higher than the normal radio
PSRs, they are not magnetars and can be argued to be the descendants
of progenitors not much more massive than that of the normal radio
PSRs. In that case, it is expected that the progenitor age
$t_{\rmn{OB}}$ will be considerably longer in comparison to that of
magnetar progenitors. An upper limit on the ages of M7 progenitors 
can be placed from the age of the Gould Belt, 
$t_{\rmn{OB}}\leq t_{\rmn{GB}}\sim 30 - 60~\rmn{Myr}$ \citep{Torraetal2000}.

From Eq. \ref{eq:age} the ages of the sample objects are proportional
to $t_{\rmn{OB}}$, thus significant uncertainty in the progenitor age
will yield erroneous ages. We estimate $t_{\rmn{OB}}$ by considering 
the total number of OB stars in the Galaxy and the supernova rate 
corresponding to type Ib/c and type II supernovae,
\begin{equation}
t_{\rmn{OB}} \approx \frac{N_{\rmn{OB,Gal}}}{\beta_{\rmn{SN}}}
\label{eq:tob}
\end{equation}
From our modeling of OB stars in the Galaxy \citep{GillHeyl2007}, we 
estimate $N_{\rmn{OB,Gal}} \sim 5.2\times10^5$, and using the supernova (SN) 
rate reported by \citet{Diehletal2006} $\beta_{\rmn{SN}} = 1.9\pm1.1$ per 
century, we find $t_{\rmn{OB}} \approx 27\pm16~\rmn{Myr}$.

\begin{table*}
\setcounter{footnote}{0}
\caption{Properties of nearby thermally emitting isolated NSs}
\begin{threeparttable}
\begin{tabular}{llllllllll}
\hline
\hline
Object & $P$ & $\dot{P}$ & $D$ & $T_{bb}$ & $N_H$ & $F_x$ & $N_{OB}$ & $t_{\rmn{stat}}\pm\Delta t_\rmn{sys}$ & $t_{\rmn{kin/dyn}}$ \\
& (s) & $(10^{-14}\mbox{ s s}^{-1})$ & (pc) & (eV) & $(10^{20}\mbox{ cm}^{-2})$ & 
 & & (Myr) & (Myr) \\
\hline
RBS 1223\tnote{1} & 10.31\tnote{2} & 11.20\tnote{2} & $\geq 525$\tnote{3} & $118\pm13$ & 
$0.5 - 2.1$ & 4.5 & $368^{+32}_{-28}$ & $0.073\pm0.005$ & $0.5-1$\tnote{25} \\
2XMM J$104608.7^\star$\tnote{4} & - & - & 2000 & $117\pm14$ & $35\pm11$ & 
0.097 & $1702^{+668}_{-477}$ & $0.089\pm0.008$ \\
RX J$1605.3 + 3249$\tnote{5} & - & - & $325 - 390$\tnote{6} & $86 - 98$ & 
$0.6 - 1.5$ & 1.15 & $181^{+414}_{-23}$ & $0.23\pm0.18$ & $0.45-3.5$\tnote{26} \\
RBS 1774\tnote{7} & 9.437\tnote{8} & $4.1\pm1.8$\tnote{9} & $390 - 430$\tnote{10} & $92^{+19}_{-15}$ 
& $4.6\pm0.2$ & 8.7 & $588^{+138}_{-75}$ & $0.28\pm0.18$ \\
RX J$0806.4 - 4123$\tnote{11} & 11.37\tnote{12} & $5.5\pm3.0$\tnote{12} & $240\pm25$\tnote{13} & 
$78\pm7$ & $2.5\pm0.9$ & 2.9 & $211^{+26}_{-27}$ & $0.41\pm0.18$ \\
RX J$0720.4 - 3125$\tnote{14} & 8.39\tnote{15} & $6.98\pm0.02$\tnote{15} & $330^{+170}_{-80}$\tnote{16} & 
$79\pm4$ & $1.3\pm0.3$ & 11.5 & $278^{+42}_{-35}$ & $0.51\pm0.18$ & $0.5-1$\tnote{17} \\
RX J$0420.0 - 5022$\tnote{18} & 3.45\tnote{19} & $2.8\pm0.3$\tnote{19} & 350\tnote{20} & $57^{+25}_{-47}$ 
& 1.7 & 0.69 & $713^{+99}_{-89}$ & $0.55\pm0.18$ \\
RX J$1856.5 - 3754$\tnote{21} & 7.06\tnote{22} & $2.97\pm0.07$\tnote{22} & $161^{+18}_{-14}$\tnote{23} 
& $57\pm1$ & $1.4\pm0.1$ & 14.6 & $604^{+256}_{-115}$ & $0.59\pm0.18$ & $\sim 0.4$\tnote{24} \\
\hline
\end{tabular}
\begin{tablenotes}[flushleft]
\setcounter{footnote}{0}
\item $F_x (10^{-12}\mbox{ erg cm}^{-2}\mbox{ s}^{-1})$ - The absorbed X-ray flux in the ROSAT energy band ($0.12 - 2.4$ keV). 
\item ${}^\star$ 2XMM J$104608.7 - 594306$
\item [1] \citet{Schwopeetal1999}
\item [2] \citet{KaplanKerkwijk2005a} 
\item [3] \citet{Posseltetal2007} 
\item [4] \citet{Piresetal2009} 
\item [5] \citet{Motchetal1999}
\item [6] \citet{Posseltetal2007}
\item [7] \citet{Zampierietal2001}
\item [8] \citet{Zaneetal2005}
\item [9] \cite{KaplanKerkwijk2009a}
\item [10] \citet{Posseltetal2007}
\item [11] \citet{Haberletal1998}
\item [12] \citet{KaplanKerkwijk2009b}
\item [13] \citet{Motchetal2008}
\item [14] \citet{Haberletal1997}
\item [15] \citet{KaplanKerkwijk2005b}
\item [16] \citet{Kaplanetal2007}
\item [17] \citet{Kaplanetal2007,Tetzlaffetal2011}
\item [18] \citet{Haberletal1999}
\item [19] \citet{KaplanKerkwijk2011}
\item [20] \citet{Posseltetal2007}
\item [21] \citet{Walteretal1996}
\item [22] \citet{KerkwijkKaplan2008}
\item [23] \citet{Kaplanetal2007}
\item [24] \citet{Mignanietal2013,Tetzlaffetal2011,Kaplanetal2002,WalterLattimer2002}
\item [25] \citet{Tetzlaffetal2010}
\item [26] \citet{Tetzlaffetal2012}
\end{tablenotes}
\end{threeparttable}
\label{tbl:m7table}
\end{table*}

Over the last few years, two new candidates have been added to the M7
group. The first object, 1RXS J$141256.0+792204$ dubbed
\textit{Calvera} \citep{Rutledgeetal2008}, was actually cataloged in
the RASS Bright Source Catalog \citep{Vogesetal1999} for having a high
X-ray to optical flux ratio $F_X/F_V > 8700$. However, its large height above the Galactic plane
($z\approx5.1$ kpc), requiring a space velocity $v_z\ga5100\rmn{ km
  s}^{-1}$, presents a challenge for its interpretation as an isolated
cooling NS like the M7 members \citep[see][for a detailed discussion]{Rutledgeetal2008}. 
Also, recent X-ray observations of Calvera done
with the XMM-Newton space telescope found unambiguous evidence for
pulsations with period $P = 59.2~\mathrm{ms}$
\citep{Zaneetal2011}. The authors of this study argued that Calvera is
most probably a CCO or a slightly recycled pulsar. The uncertainty in
its nature \citep[see][]{halpern2011} doesn't warrant inclusion into
our sample of radio-quiet isolated NSs. The second object 2XMM
J$104608.7 - 594306$ \citep{Piresetal2009}, discovered serendipitously
in an XMM-Newton pointed observation of the Carina Nebula hosting the
binary system Eta Carinae, appears to be a promising candidate (see
Table \ref{tbl:m7table} for properties). This object was not detected
in the RASS due to its larger distance ($\approx 2.3$ kpc, based on
its association to the Carina nebula) and higher neutral hydrogen
absorption column density ($N_H = 3.5\pm1.1\times10^{21}\mbox{
  cm}^{-2}$). Therefore, the accessible volume $V_{\rmn{max}}$ is the
ROSAT surveyed volume plus the additional volume probed by the
XMM-Newton's pointed observation.

\begin{figure}
\centering
\includegraphics[width=0.5\textwidth]{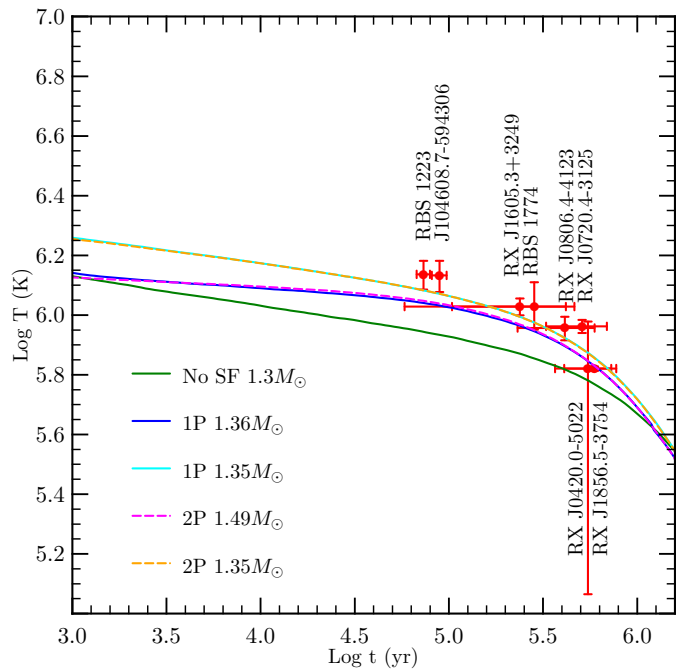}
\caption{Cooling curve of nearby thermally emitting isolated NSs obtained from 
model temperatures and statistical ages. The errors in the ages reflect only the 
systematic error (see Sec. \ref{sec:errors} for details). We plot 
sample theoretical cooling curves from \citet{YakovlevPethick2004} for comparison.}
\label{fig:m7cool}
\end{figure}

In Table \ref{tbl:m7table}, we provide all the relevant data on the
sample objects including the spectral fit parameters that were used to
simulate the RASS to obtain $V_{\mathrm{max}}$. We take the calculated
ages and plot them against the effective temperatures observed at
infinity in Fig. \ref{fig:m7cool}. The errorbars on the ages
correspond to the maximum of the difference in ages obtained due to
uncertainties in $T_{bb}$, $N_H$ (these two parameters are
covariant), and the distance. The blackbody temperature $T_{bb}$ is 
obtained by fitting a blackbody spectrum to that observed from the source. 
The temperature, thus, corresponds to the color temperature of the object 
and is an overestimation of the effective temperature $T_{eff}$ due to strong 
energy dependence of the free-free and bound-free opacities of the photosphere 
\citep[see for e.g.][]{Lloydetal2003}. The effective temperature is obtained from 
$T_{bb}$ using a color correction factor $f_c = T_{bb}/T_{eff}$ where 
$1\la f_c\la 1.8$ \citep[e.g.][]{Ozel2013}. For comparison, we also plot some
cooling curves from \citet{YakovlevPethick2004}, where the
non-superfluid (No SF) model for a $1.3 M_\odot$ cannot explain the
data. Other model curves show NS cooling behavior if proton
superfluidity in the core is taken into account 
\citep[see][for more details on the 1P and 2P models]{YakovlevPethick2004}.
\subsection{\label{sec:errors}Statistical Ages Vs The True Ages}
The ages of isolated NSs have been estimated using different methods, namely from the 
spin-down law, cooling models, and kinematics. The method we propose in this study to 
estimate the true ages of these objects has only been applied, in its original form, 
to estimate the ages of white dwarfs from their cumulative luminosity
function in globular cluster \citep[see for e.g.][]{Goldsburyetal2012}. The method 
itself is purely a statistical one, for which the underlying assumption is that the 
objects in the sample follow a Poisson distribution \citep{Felton1976} with a constant production rate. 
The important question to ask here is 
how good of an estimate of the true age is the statistical age. What is 
the inherent statistical and other errors associated to this method of predicting ages? 

\subsubsection{Statistical Error}
Consider 
the youngest object, RBS 1223, which is also the hottest among the eight isolated NSs. 
The kinematic age of RBS 1223 has been 
found based on its association to possible OB associations and young star clusters to 
be $\sim0.5-1$ Myr \citep{Tetzlaffetal2010}, 
with large uncertainties. The statistical age of RBS 1223 that we find 
in our study, given that the sample only contains eight such objects, 
is much smaller $\sim 0.073\pm0.005$ Myr, with uncertainties 
corresponding to systematic errors. The statistical error is of course 
much larger than the systematic one. Since the discovery of an 
object in a given volume follows Poisson statistics, the relative 
error scales as $1/\sqrt{N}$ where $N$ is the sample size with ages 
$t_i \leq t_N$. Therefore, the error in the age of the first object 
is $\pm t_1$ where $t_1$ is its statistical age. Likewise, the error 
in the age of the eighth object is $\pm t_8/\sqrt{8}$. Evidently, 
one needs a much larger sample to reduce the statistical errors 
to that comparable to the systematic errors.
\subsubsection{Other Errors}
The grand assumption made in this work is that all M7 members are essentially 
the same object but with different ages. Thus, they all follow one cooling 
curve. That may not be the complete story. These objects may have different 
masses, surface composition, magnetic fields, and they also may be located 
with a line of sight with an inhomogeneous HI column density. All of these 
parameters affect the inferred temperatures and observability of these objects, and 
hence their statistical ages \citep[see for e.g. the review by][]{YakovlevPethick2004}, 
as the objects were ordered from hottest to coldest in Eq. \ref{eq:age}. Even in a larger sample 
of isolated NSs, the dominant errors come from systematics and the uncertainty in the progenitor age $t_\rmn{OB}$. 
In the case of RX J1856.5-3754, 
the statistical age with its systematic error is $t_\rmn{stat}=0.59\pm0.18~\rmn{Myr}$. 
Including the statistical error $\Delta t_\rmn{stat} = 0.59/\sqrt{8}~\rmn{Myr}$, since it's the 
$8^\rmn{th}$ object in the temperature ordered list, and the error due to the uncertainty 
$\Delta t_\rmn{OB}=16~\rmn{Myr}$, the total error is $\Delta t_\rmn{tot} = 0.44~\rmn{Myr}$. 
Systematic errors can be reduced by having better estimates of distances, senstive surveys that 
yield accurate surface temperatures, and an improved model of the HI column density. Accurate 
determination of these parameters is crucial to obtain accurate estimates of $V_\rmn{max}$.

For RX J0720.4-3125, we find that the total error is $\Delta t_\rmn{tot} = 0.41~\rmn{Myr}$. Although 
we find this object to be younger than RX J1856.5-3754, the large uncertainties in the statistical 
ages of both objects make it hard to distinguish which is younger. The kinematic ages of these 
objects tell a different story, where RX J0720.4-3125 is older than RX J1856.5-3754. This 
discrepancy results from the ordering of objects based on their model temperatures. If 
both objects are indeed very similar in their mass and surface composition, and have 
cooled not much differently, then the statistical ages would agree with RX J0720.4-3125 
being younger.
\section{Discussion}
The ages of isolated NSs are primarily important for constraining their 
inner structure. In addition, they can be useful for accurately 
determining the birthplace of the object, if its proper motion 
is known. Similarly, if the SNR-NS association has been made, 
then the age of the object can reveal its space velocity and the 
kinematics of the SNR. The knowledge of ages of such objects is also 
useful for population synthesis models which rely on the spatial and 
velocity distributions, and the birthrates of NSs. In this study, 
we propose a statistical method to estimate the true ages of the 
ROSAT discovered sample comprising the M7. We then use the age estimates 
along with the derived spectral temperatures to compare the data 
with some cooling models. The strength of this technique, as 
discussed earlier, lies in obtaining a larger sample of \textit{coolers}. 
With only eight objects, the statistical error is indubitably much larger.

We have used the statistical ages and model temperatures of the nearby thermally 
emitting isolated NSs to derive a cooling curve, under the assumption that all 
of the objects in the sample are similar in their mass and surface composition, and 
have also cooled in a similar fashion. Due to the marginal size of the sample, with 
large statistical errors, it is not yet meaningful to differentiate between the 
different cooling models. However, this whole exercise shows that ages of such 
objects can be inferred statistically and provides another way of determining 
accurate ages in addition to ages obtained from kinematics and/or spin-down.

The statistics can be improved by locating more of these objects in the disk of the Galaxy. 
In a recent population synthesis study, \citet{Posseltetal2010} find that young isolated NSs, 
that are both hot and bright, with ROSAT count rates below $0.1\mbox{ cts s}^{-1}$ (the ROSAT bright source catalog had a limiting count rate of $0.05\mbox{ cts s}^{-1}$) should be located in OB associations beyond the Gould belt. They 
also remark on the possibility of finding new isolated cooling NSs by
conducting yet another careful search of such objects in the RASS and 
the XMM-Newton Slew Survey \citep{Esquejetal2006}. However, they note that ROSAT 
observations are incapable at locating the isolated sources with sufficient 
spatial accuracy, such that many optical counterparts can be found in its large 
positional error circle. On the other hand, although XMM-Newton is much more sensitive, 
albeit with strong inhomogeneities, and can probe deeper into the Galactic plane, 
the slew survey only covers $15\%$ of the sky currently. Searching the RASS for new 
isolated NSs may appear to be a promising avenue \citep[e.g.][]{Turneretal2010}. 
What is needed at the moment is another all-sky survey 
that is able to surpass ROSAT in both sensitivity and positional 
accuracy. In that regard, the upcoming eROSITA mission \citep{Cappellutietal2011} shows 
a lot of promise, and its planned launch in 2014 makes it very timely. The X-ray 
instrument eROSITA will be part of the Russian Spectrum-Roentgen-Gamma (SRG) satellite, 
equipped with seven Wolter-I telescope modules with an advanced version of the XMM-Newton 
pnCCD camera at its prime focus. The telescope will operate with 
an energy range of $0.5 - 10$ keV, a field of view (FOV) of $1.03^\circ$, an angular 
resolution of $28''$ averaged over the FOV, and a limiting flux of 
$\sim10^{-14}\mbox{ erg cm}^{-3}$ in the $0.5 - 2$ keV 
energy range and $\sim3\times10^{-13}\mbox{erg cm}^{-3}$ in the $2 - 10$ 
keV energy range. The all-sky survey will reach sensitivities that are $\sim 30$ times 
that of the RASS where the entire sky will scanned over a period of four years.
\section{acknowledgements}
We would like to thank the referee for significantly 
improving the quality of this work. The Natural
Sciences and Engineering Research Council of Canada,
Canadian Foundation for Innovation and the British
Columbia Knowledge Development Fund supported this
work. Correspondence and requests for materials should
be addressed to J.S.H. (heyl@phas.ubc.ca). This research
has made use of NASAs Astrophysics Data System 
Bibliographic Services
\bibliographystyle{mn2e}

\begin{thebibliography}{}

\bibitem[\protect\citeauthoryear{{An}, {Kaspi}, {Archibald} \& {Cumming}}{{An}
  et~al.}{2013}]{Anetal2013}
{An} H.,  {Kaspi} V.~M.,  {Archibald} R.,    {Cumming} A.,  2013, \apj, 763, 82

\bibitem[\protect\citeauthoryear{{Aschenbach}, {Egger} \&
  {Tr{\"u}mper}}{{Aschenbach} et~al.}{1995}]{Aschenbachetal1995}
{Aschenbach} B.,  {Egger} R.,    {Tr{\"u}mper} J.,  1995, \nat, 373, 587

\bibitem[\protect\citeauthoryear{{Avni} \& {Bahcall}}{{Avni} \&
  {Bahcall}}{1980}]{AvniBahcall1980}
{Avni} Y.,  {Bahcall} J.~N.,  1980, \apj, 235, 694

\bibitem[\protect\citeauthoryear{{Bahcall} \& {Soneira}}{{Bahcall} \&
  {Soneira}}{1980}]{BahcallSoneira1980}
{Bahcall} J.~N.,  {Soneira} R.~M.,  1980, ApJ, Supplement, 44, 73

\bibitem[\protect\citeauthoryear{{Becker}}{{Becker}}{2009}]{Becker2009}
{Becker} W.,  ed. 2009, {Neutron Stars and Pulsars} Vol.~357 of Astrophysics
  and Space Science Library

\bibitem[\protect\citeauthoryear{{Becker} \& {Pavlov}}{{Becker} \&
  {Pavlov}}{2002}]{BeckerPavlov2002}
{Becker} W.,  {Pavlov} G.~G.,  2002, ArXiv Astrophysics e-prints

\bibitem[\protect\citeauthoryear{{Becker}, {Prinz}, {Winkler} \&
  {Petre}}{{Becker} et~al.}{2012}]{Beckeretal2012}
{Becker} W.,  {Prinz} T.,  {Winkler} P.~F.,    {Petre} R.,  2012, \apj, 755,
  141

\bibitem[\protect\citeauthoryear{{Binney} \& {Merrifield}}{{Binney} \&
  {Merrifield}}{1998}]{BinneyMerrifield1998}
{Binney} J.,  {Merrifield} M.,  1998, {Galactic astronomy}.
Princeton University Press

\bibitem[\protect\citeauthoryear{{Blazek}, {Gaensler}, {Chatterjee}, {van der
  Swaluw}, {Camilo} \& {Stappers}}{{Blazek} et~al.}{2006}]{Blazeketal2006}
{Blazek} J.~A.,  {Gaensler} B.~M.,  {Chatterjee} S.,  {van der Swaluw} E.,
  {Camilo} F.,    {Stappers} B.~W.,  2006, \apj, 652, 1523

\bibitem[\protect\citeauthoryear{{Cappelluti}, {Predehl}, {B{\"o}hringer},
  {Brunner}, {Brusa}, {Burwitz}, {Churazov}, {Dennerl}, {Finoguenov},
  {Freyberg}, {Friedrich}, {Hasinger}, {Kenziorra} \&
  {Kreykenbohm}}{{Cappelluti} et~al.}{2011}]{Cappellutietal2011}
{Cappelluti} N.,  {Predehl} P.,  {B{\"o}hringer} H.,  {Brunner} H.,  {Brusa}
  M.,  {Burwitz} V.,  {Churazov} E.,  {Dennerl} K.,  {Finoguenov} A.,
  {Freyberg} M.,  {Friedrich} P.,  {Hasinger} G.,  {Kenziorra} E.,
  {Kreykenbohm} 2011, Memorie della Societa Astronomica Italiana Supplementi,
  17, 159

\bibitem[\protect\citeauthoryear{{Caswell}, {Kesteven}, {Stewart}, {Milne} \&
  {Haynes}}{{Caswell} et~al.}{1992}]{Caswelletal1992}
{Caswell} J.~L.,  {Kesteven} M.~J.,  {Stewart} R.~T.,  {Milne} D.~K.,
  {Haynes} R.~F.,  1992, \apjl, 399, L151

\bibitem[\protect\citeauthoryear{{Clifton} \& {Lyne}}{{Clifton} \&
  {Lyne}}{1986}]{CliftonLyne1986}
{Clifton} T.~R.,  {Lyne} A.~G.,  1986, \nat, 320, 43

\bibitem[\protect\citeauthoryear{{Colpi}, {Geppert} \& {Page}}{{Colpi}
  et~al.}{2000}]{Colpietal2000}
{Colpi} M.,  {Geppert} U.,    {Page} D.,  2000, \apjl, 529, L29

\bibitem[\protect\citeauthoryear{{Corbel}, {Chapuis}, {Dame} \&
  {Durouchoux}}{{Corbel} et~al.}{1999}]{Corbeletal1999}
{Corbel} S.,  {Chapuis} C.,  {Dame} T.~M.,    {Durouchoux} P.,  1999, \apjl,
  526, L29

\bibitem[\protect\citeauthoryear{{Davies}, {Figer}, {Kudritzki}, {Trombley},
  {Kouveliotou} \& {Wachter}}{{Davies} et~al.}{2009}]{Daviesetal2009}
{Davies} B.,  {Figer} D.~F.,  {Kudritzki} R.-P.,  {Trombley} C.,  {Kouveliotou}
  C.,    {Wachter} S.,  2009, \apj, 707, 844

\bibitem[\protect\citeauthoryear{{Demorest}, {Pennucci}, {Ransom}, {Roberts} \&
  {Hessels}}{{Demorest} et~al.}{2010}]{Demorestetal2010}
{Demorest} P.~B.,  {Pennucci} T.,  {Ransom} S.~M.,  {Roberts} M.~S.~E.,
  {Hessels} J.~W.~T.,  2010, \nat, 467, 1081

\bibitem[\protect\citeauthoryear{{Dermer} \& {Powale}}{{Dermer} \&
  {Powale}}{2013}]{DermerPowale2013}
{Dermer} C.~D.,  {Powale} G.,  2013, \aap, 553, A34

\bibitem[\protect\citeauthoryear{{Dewey}, {Taylor}, {Weisberg} \&
  {Stokes}}{{Dewey} et~al.}{1985}]{Deweyetal1985}
{Dewey} R.~J.,  {Taylor} J.~H.,  {Weisberg} J.~M.,    {Stokes} G.~H.,  1985,
  \apjl, 294, L25

\bibitem[\protect\citeauthoryear{{Dib}, {Kaspi} \& {Gavriil}}{{Dib}
  et~al.}{2008}]{Dibetal2008}
{Dib} R.,  {Kaspi} V.~M.,    {Gavriil} F.~P.,  2008, \apj, 673, 1044

\bibitem[\protect\citeauthoryear{{Dib}, {Kaspi} \& {Gavriil}}{{Dib}
  et~al.}{2009}]{Dibetal2009}
{Dib} R.,  {Kaspi} V.~M.,    {Gavriil} F.~P.,  2009, \apj, 702, 614

\bibitem[\protect\citeauthoryear{{Diehl}, {Halloin}, {Kretschmer}, {Lichti},
  {Sch{\"o}nfelder}, {Strong}, {von Kienlin}, {Wang}, {Jean}, {Kn{\"o}dlseder},
  {Roques}, {Weidenspointner}, {Schanne}, {Hartmann}, {Winkler} \&
  {Wunderer}}{{Diehl} et~al.}{2006}]{Diehletal2006}
{Diehl} R.,  {Halloin} H.,  {Kretschmer} K.,  {Lichti} G.~G.,
  {Sch{\"o}nfelder} V.,  {Strong} A.~W.,  {von Kienlin} A.,  {Wang} W.,  {Jean}
  P.,  {Kn{\"o}dlseder} J.,  {Roques} J.-P.,  {Weidenspointner} G.,  {Schanne}
  S.,  {Hartmann} D.~H.,  {Winkler} C.,    {Wunderer} C.,  2006, \nat, 439, 45

\bibitem[\protect\citeauthoryear{{Dodson}, {McCulloch} \& {Lewis}}{{Dodson}
  et~al.}{2002}]{Dodsonetal2002}
{Dodson} R.~G.,  {McCulloch} P.~M.,    {Lewis} D.~R.,  2002, \apjl, 564, L85

\bibitem[\protect\citeauthoryear{{Esposito}, {Burgay}, {Possenti}, {Turolla},
  {Zane}, {de Luca}, {Tiengo}, {Israel}, {Mattana}, {Mereghetti}, {Bailes},
  {Romano}, {G{\"o}tz} \& {Rea}}{{Esposito} et~al.}{2009}]{Espositoetal2009a}
{Esposito} P.,  {Burgay} M.,  {Possenti} A.,  {Turolla} R.,  {Zane} S.,  {de
  Luca} A.,  {Tiengo} A.,  {Israel} G.~L.,  {Mattana} F.,  {Mereghetti} S.,
  {Bailes} M.,  {Romano} P.,  {G{\"o}tz} D.,    {Rea} N.,  2009, \mnras, 399,
  L44

\bibitem[\protect\citeauthoryear{{Esposito}, {Tiengo}, {Mereghetti}, {Israel},
  {De Luca}, {G{\"o}tz}, {Rea}, {Turolla} \& {Zane}}{{Esposito}
  et~al.}{2009}]{Espositoetal2009}
{Esposito} P.,  {Tiengo} A.,  {Mereghetti} S.,  {Israel} G.~L.,  {De Luca} A.,
  {G{\"o}tz} D.,  {Rea} N.,  {Turolla} R.,    {Zane} S.,  2009, \apjl, 690,
  L105

\bibitem[\protect\citeauthoryear{{Esquej}, {Altieri}, {Bermejo}, {Freyberg},
  {Lazaro}, {Read} \& {Saxton}}{{Esquej} et~al.}{2006}]{Esquejetal2006}
{Esquej} M.~P.,  {Altieri} B.,  {Bermejo} D.,  {Freyberg} M.~J.,  {Lazaro} V.,
  {Read} A.~M.,    {Saxton} R.~D.,  2006, in {A.~Wilson} ed., The X-ray
  Universe 2005 Vol.~604 of ESA Special Publication, {The XMM-Newton Slew
  Survey: Towards the XMMSL1 Catalogue}.
p.~965

\bibitem[\protect\citeauthoryear{{Felten}}{{Felten}}{1976}]{Felton1976}
{Felten} J.~E.,  1976, \apj, 207, 700

\bibitem[\protect\citeauthoryear{{Figer}, {Najarro}, {Geballe}, {Blum} \&
  {Kudritzki}}{{Figer} et~al.}{2005}]{Figeretal2005}
{Figer} D.~F.,  {Najarro} F.,  {Geballe} T.~R.,  {Blum} R.~D.,    {Kudritzki}
  R.~P.,  2005, \apjl, 622, L49

\bibitem[\protect\citeauthoryear{{Finley} \& {Oegelman}}{{Finley} \&
  {Oegelman}}{1994}]{FinleyOegelman1994}
{Finley} J.~P.,  {Oegelman} H.,  1994, \apjl, 434, L25

\bibitem[\protect\citeauthoryear{{Foster} \& {Routledge}}{{Foster} \&
  {Routledge}}{2003}]{FosterRoutledge2003}
{Foster} T.,  {Routledge} D.,  2003, ApJ, 598, 1005

\bibitem[\protect\citeauthoryear{{Frail}, {Goss} \& {Whiteoak}}{{Frail}
  et~al.}{1994}]{Frailetal1994}
{Frail} D.~A.,  {Goss} W.~M.,    {Whiteoak} J.~B.~Z.,  1994, \apj, 437, 781

\bibitem[\protect\citeauthoryear{{Gaensler}, {McClure-Griffiths}, {Oey},
  {Haverkorn}, {Dickey} \& {Green}}{{Gaensler} et~al.}{2005}]{Gaensleretal2005}
{Gaensler} B.~M.,  {McClure-Griffiths} N.~M.,  {Oey} M.~S.,  {Haverkorn} M.,
  {Dickey} J.~M.,    {Green} A.~J.,  2005, \apjl, 620, L95

\bibitem[\protect\citeauthoryear{{Gavriil} \& {Kaspi}}{{Gavriil} \&
  {Kaspi}}{2002}]{GavriilKaspi2002}
{Gavriil} F.~P.,  {Kaspi} V.~M.,  2002, \apj, 567, 1067

\bibitem[\protect\citeauthoryear{{Gill} \& {Heyl}}{{Gill} \&
  {Heyl}}{2007}]{GillHeyl2007}
{Gill} R.,  {Heyl} J.,  2007, \mnras, 381, 52

\bibitem[\protect\citeauthoryear{{Goldreich} \& {Reisenegger}}{{Goldreich} \&
  {Reisenegger}}{1992}]{GoldreichReisenegger1992}
{Goldreich} P.,  {Reisenegger} A.,  1992, \apj, 395, 250

\bibitem[\protect\citeauthoryear{{Goldsbury}, {Heyl}, {Richer}, {Bergeron},
  {Dotter}, {Kalirai}, {MacDonald}, {Rich}, {Stetson}, {Tremblay} \&
  {Woodley}}{{Goldsbury} et~al.}{2012}]{Goldsburyetal2012}
{Goldsbury} R.,  {Heyl} J.,  {Richer} H.~B.,  {Bergeron} P.,  {Dotter} A.,
  {Kalirai} J.~S.,  {MacDonald} J.,  {Rich} R.~M.,  {Stetson} P.~B.,
  {Tremblay} P.-E.,    {Woodley} K.~A.,  2012, \apj, 760, 78

\bibitem[\protect\citeauthoryear{{Gotthelf}, {Halpern} \& {Alford}}{{Gotthelf}
  et~al.}{2013}]{Gotthelfetal2013}
{Gotthelf} E.~V.,  {Halpern} J.~P.,    {Alford} J.,  2013, \apj, 765, 58

\bibitem[\protect\citeauthoryear{{Haberl}}{{Haberl}}{2007}]{Haberl2007}
{Haberl} F.,  2007, \apss, 308, 181

\bibitem[\protect\citeauthoryear{{Haberl}, {Motch}, {Buckley}, {Zickgraf} \&
  {Pietsch}}{{Haberl} et~al.}{1997}]{Haberletal1997}
{Haberl} F.,  {Motch} C.,  {Buckley} D.~A.~H.,  {Zickgraf} F.-J.,    {Pietsch}
  W.,  1997, \aap, 326, 662

\bibitem[\protect\citeauthoryear{{Haberl}, {Motch} \& {Pietsch}}{{Haberl}
  et~al.}{1998}]{Haberletal1998}
{Haberl} F.,  {Motch} C.,    {Pietsch} W.,  1998, Astronomische Nachrichten,
  319, 97

\bibitem[\protect\citeauthoryear{{Haberl}, {Pietsch} \& {Motch}}{{Haberl}
  et~al.}{1999}]{Haberletal1999}
{Haberl} F.,  {Pietsch} W.,    {Motch} C.,  1999, \aap, 351, L53

\bibitem[\protect\citeauthoryear{{Halpern}}{{Halpern}}{2011}]{halpern2011}
{Halpern} J.~P.,  2011, \apjl, 736, L3

\bibitem[\protect\citeauthoryear{{Halpern} \& {Gotthelf}}{{Halpern} \&
  {Gotthelf}}{2010a}]{HalpernGotthelf2010}
{Halpern} J.~P.,  {Gotthelf} E.~V.,  2010a, \apj, 709, 436

\bibitem[\protect\citeauthoryear{{Halpern} \& {Gotthelf}}{{Halpern} \&
  {Gotthelf}}{2010b}]{HalpernGotthelf2010a}
{Halpern} J.~P.,  {Gotthelf} E.~V.,  2010b, \apj, 710, 941

\bibitem[\protect\citeauthoryear{{Hartwick} \& {Schade}}{{Hartwick} \&
  {Schade}}{1990}]{HartwickSchade1990}
{Hartwick} F.~D.~A.,  {Schade} D.,  1990, \araa, 28, 437

\bibitem[\protect\citeauthoryear{{Hester}}{{Hester}}{2008}]{Hester2008}
{Hester} J.~J.,  2008, \araa, 46, 127

\bibitem[\protect\citeauthoryear{{Heyl} \& {Hernquist}}{{Heyl} \&
  {Hernquist}}{1998}]{HeylHernquist1998}
{Heyl} J.~S.,  {Hernquist} L.,  1998, \mnras, 297, L69

\bibitem[\protect\citeauthoryear{{Heyl} \& {Kulkarni}}{{Heyl} \&
  {Kulkarni}}{1998}]{HeylKulkarni1998}
{Heyl} J.~S.,  {Kulkarni} S.~R.,  1998, \apjl, 506, L61

\bibitem[\protect\citeauthoryear{{Hobbs}, {Lyne}, {Kramer}, {Martin} \&
  {Jordan}}{{Hobbs} et~al.}{2004}]{Hobbsetal2004}
{Hobbs} G.,  {Lyne} A.~G.,  {Kramer} M.,  {Martin} C.~E.,    {Jordan} C.,
  2004, \mnras, 353, 1311

\bibitem[\protect\citeauthoryear{{Huchra} \& {Sargent}}{{Huchra} \&
  {Sargent}}{1973}]{HuchraSargent1973}
{Huchra} J.,  {Sargent} W.~L.~W.,  1973, \apj, 186, 433

\bibitem[\protect\citeauthoryear{{Hulse} \& {Taylor}}{{Hulse} \&
  {Taylor}}{1975}]{HulseTaylor1975}
{Hulse} R.~A.,  {Taylor} J.~H.,  1975, \apjl, 201, L55

\bibitem[\protect\citeauthoryear{{H{\"u}nsch}, {Schmitt}, {Sterzik} \&
  {Voges}}{{H{\"u}nsch} et~al.}{1999}]{Hunschetal1999}
{H{\"u}nsch} M.,  {Schmitt} J.~H.~M.~M.,  {Sterzik} M.~F.,    {Voges} W.,
  1999, \aaps, 135, 319

\bibitem[\protect\citeauthoryear{{Johnston}, {Lyne}, {Manchester}, {Kniffen},
  {D'Amico}, {Lim} \& {Ashworth}}{{Johnston} et~al.}{1992}]{Johnstonetal1992}
{Johnston} S.,  {Lyne} A.~G.,  {Manchester} R.~N.,  {Kniffen} D.~A.,  {D'Amico}
  N.,  {Lim} J.,    {Ashworth} M.,  1992, \mnras, 255, 401

\bibitem[\protect\citeauthoryear{{Kaplan}, {Kulkarni}, {van Kerkwijk} \&
  {Marshall}}{{Kaplan} et~al.}{2002}]{Kaplanetal2002b}
{Kaplan} D.~L.,  {Kulkarni} S.~R.,  {van Kerkwijk} M.~H.,    {Marshall} H.~L.,
  2002, \apjl, 570, L79

\bibitem[\protect\citeauthoryear{{Kaplan} \& {van Kerkwijk}}{{Kaplan} \& {van
  Kerkwijk}}{2005a}]{KaplanKerkwijk2005b}
{Kaplan} D.~L.,  {van Kerkwijk} M.~H.,  2005a, \apjl, 628, L45

\bibitem[\protect\citeauthoryear{{Kaplan} \& {van Kerkwijk}}{{Kaplan} \& {van
  Kerkwijk}}{2005b}]{KaplanKerkwijk2005a}
{Kaplan} D.~L.,  {van Kerkwijk} M.~H.,  2005b, \apjl, 635, L65

\bibitem[\protect\citeauthoryear{{Kaplan} \& {van Kerkwijk}}{{Kaplan} \& {van
  Kerkwijk}}{2009a}]{KaplanKerkwijk2009b}
{Kaplan} D.~L.,  {van Kerkwijk} M.~H.,  2009a, \apj, 705, 798

\bibitem[\protect\citeauthoryear{{Kaplan} \& {van Kerkwijk}}{{Kaplan} \& {van
  Kerkwijk}}{2009b}]{KaplanKerkwijk2009a}
{Kaplan} D.~L.,  {van Kerkwijk} M.~H.,  2009b, \apjl, 692, L62

\bibitem[\protect\citeauthoryear{{Kaplan} \& {van Kerkwijk}}{{Kaplan} \& {van
  Kerkwijk}}{2011}]{KaplanKerkwijk2011}
{Kaplan} D.~L.,  {van Kerkwijk} M.~H.,  2011, \apjl, 740, L30

\bibitem[\protect\citeauthoryear{{Kaplan}, {van Kerkwijk} \&
  {Anderson}}{{Kaplan} et~al.}{2002}]{Kaplanetal2002}
{Kaplan} D.~L.,  {van Kerkwijk} M.~H.,    {Anderson} J.,  2002, \apj, 571, 447

\bibitem[\protect\citeauthoryear{{Kaplan}, {van Kerkwijk} \&
  {Anderson}}{{Kaplan} et~al.}{2007}]{Kaplanetal2007}
{Kaplan} D.~L.,  {van Kerkwijk} M.~H.,    {Anderson} J.,  2007, \apj, 660, 1428

\bibitem[\protect\citeauthoryear{{Kaspi}, {Manchester}, {Siegman}, {Johnston}
  \& {Lyne}}{{Kaspi} et~al.}{1994}]{Kaspietal1994}
{Kaspi} V.~M.,  {Manchester} R.~N.,  {Siegman} B.,  {Johnston} S.,    {Lyne}
  A.~G.,  1994, \apjl, 422, L83

\bibitem[\protect\citeauthoryear{{Kovalenko}}{{Kovalenko}}{1989}]{Kovalenko1989}
{Kovalenko} A.~V.,  1989, Soviet Astronomy Letters, 15, 144

\bibitem[\protect\citeauthoryear{{Large}, {Vaughan} \& {Mills}}{{Large}
  et~al.}{1968}]{Largeetal1968}
{Large} M.~I.,  {Vaughan} A.~E.,    {Mills} B.~Y.,  1968, \nat, 220, 340

\bibitem[\protect\citeauthoryear{{Lloyd}, {Hernquist} \& {Heyl}}{{Lloyd}
  et~al.}{2003}]{Lloydetal2003}
{Lloyd} D.~A.,  {Hernquist} L.,    {Heyl} J.~S.,  2003, \apj, 593, 1024

\bibitem[\protect\citeauthoryear{{Lyne} \& {Graham-Smith}}{{Lyne} \&
  {Graham-Smith}}{2006}]{lynesmith2006}
{Lyne} A.~G.,  {Graham-Smith} F.,  2006, Pulsar Astronomy.
Cambridge University Press

\bibitem[\protect\citeauthoryear{{Lyne}, {Pritchard}, {Graham-Smith} \&
  {Camilo}}{{Lyne} et~al.}{1996}]{Lyneetal1996}
{Lyne} A.~G.,  {Pritchard} R.~S.,  {Graham-Smith} F.,    {Camilo} F.,  1996,
  \nat, 381, 497

\bibitem[\protect\citeauthoryear{{Manchester}, {Damico} \&
  {Tuohy}}{{Manchester} et~al.}{1985}]{Manchesteretal1985}
{Manchester} R.~N.,  {Damico} N.,    {Tuohy} I.~R.,  1985, \mnras, 212, 975

\bibitem[\protect\citeauthoryear{{Manchester}, {Hobbs}, {Teoh} \&
  {Hobbs}}{{Manchester} et~al.}{2005}]{Manchesteretal2005}
{Manchester} R.~N.,  {Hobbs} G.~B.,  {Teoh} A.,    {Hobbs} M.,  2005, VizieR
  Online Data Catalog, 7245, 0

\bibitem[\protect\citeauthoryear{{Manchester}, {Tuohy} \&
  {Damico}}{{Manchester} et~al.}{1982}]{Manchesteretal1982}
{Manchester} R.~N.,  {Tuohy} I.~R.,    {Damico} N.,  1982, \apjl, 262, L31

\bibitem[\protect\citeauthoryear{{Mereghetti}}{{Mereghetti}}{2011}]{Mereghetti2011}
{Mereghetti} S.,  2011, in {Torres} D.~F.,  {Rea} N.,  eds, High-Energy
  Emission from Pulsars and their Systems {X-ray emission from isolated neutron
  stars}.
p.~345

\bibitem[\protect\citeauthoryear{{Mereghetti}, {Esposito}, {Tiengo}, {Zane},
  {Turolla}, {Stella}, {Israel}, {G{\"o}tz} \& {Feroci}}{{Mereghetti}
  et~al.}{2006}]{Mereghettietal2006}
{Mereghetti} S.,  {Esposito} P.,  {Tiengo} A.,  {Zane} S.,  {Turolla} R.,
  {Stella} L.,  {Israel} G.~L.,  {G{\"o}tz} D.,    {Feroci} M.,  2006, \apj,
  653, 1423

\bibitem[\protect\citeauthoryear{{Mignani}, {Vande Putte}, {Cropper},
  {Turolla}, {Zane}, {Pellizza}, {Bignone}, {Sartore} \& {Treves}}{{Mignani}
  et~al.}{2013}]{Mignanietal2013}
{Mignani} R.~P.,  {Vande Putte} D.,  {Cropper} M.,  {Turolla} R.,  {Zane} S.,
  {Pellizza} L.~J.,  {Bignone} L.~A.,  {Sartore} N.,    {Treves} A.,  2013,
  \mnras, 429, 3517

\bibitem[\protect\citeauthoryear{{Motch}, {Haberl}, {Zickgraf}, {Hasinger} \&
  {Schwope}}{{Motch} et~al.}{1999}]{Motchetal1999}
{Motch} C.,  {Haberl} F.,  {Zickgraf} F.-J.,  {Hasinger} G.,    {Schwope}
  A.~D.,  1999, \aap, 351, 177

\bibitem[\protect\citeauthoryear{{Motch}, {Pires}, {Haberl}, {Schwope} \&
  {Zavlin}}{{Motch} et~al.}{2008}]{Motchetal2008}
{Motch} C.,  {Pires} A.~M.,  {Haberl} F.,  {Schwope} A.,    {Zavlin} V.~E.,
  2008, in {C.~Bassa, Z.~Wang, A.~Cumming, \& V.~M.~Kaspi} ed., 40 Years of
  Pulsars: Millisecond Pulsars, Magnetars and More Vol.~983 of American
  Institute of Physics Conference Series, {Proper motions of ROSAT discovered
  isolated neutron stars measured with Chandra: First X-ray measurement of the
  large proper motion of RX J1308.6+2127/RBS 1223}.
pp 354--356

\bibitem[\protect\citeauthoryear{{Motch}, {Pires}, {Haberl}, {Schwope} \&
  {Zavlin}}{{Motch} et~al.}{2009}]{Motchetal2009}
{Motch} C.,  {Pires} A.~M.,  {Haberl} F.,  {Schwope} A.,    {Zavlin} V.~E.,
  2009, \aap, 497, 423

\bibitem[\protect\citeauthoryear{{Motch}, {Sekiguchi}, {Haberl}, {Zavlin},
  {Schwope} \& {Pakull}}{{Motch} et~al.}{2005}]{Motchetal2005}
{Motch} C.,  {Sekiguchi} K.,  {Haberl} F.,  {Zavlin} V.~E.,  {Schwope} A.,
  {Pakull} M.~W.,  2005, \aap, 429, 257

\bibitem[\protect\citeauthoryear{{Muno}, {Clark}, {Crowther}, {Dougherty}, {de
  Grijs}, {Law}, {McMillan}, {Morris}, {Negueruela}, {Pooley}, {Portegies
  Zwart} \& {Yusef-Zadeh}}{{Muno} et~al.}{2006}]{Munoetal2006}
{Muno} M.~P.,  {Clark} J.~S.,  {Crowther} P.~A.,  {Dougherty} S.~M.,  {de
  Grijs} R.,  {Law} C.,  {McMillan} S.~L.~W.,  {Morris} M.~R.,  {Negueruela}
  I.,  {Pooley} D.,  {Portegies Zwart} S.,    {Yusef-Zadeh} F.,  2006, \apjl,
  636, L41

\bibitem[\protect\citeauthoryear{{Nakagawa}, {Mihara}, {Yoshida}, {Yamaoka},
  {Sugita}, {Murakami}, {Yonetoku}, {Suzuki}, {Nakajima}, {Tashiro} \&
  {Nakazawa}}{{Nakagawa} et~al.}{2009}]{Nakagawaetal2009}
{Nakagawa} Y.~E.,  {Mihara} T.,  {Yoshida} A.,  {Yamaoka} K.,  {Sugita} S.,
  {Murakami} T.,  {Yonetoku} D.,  {Suzuki} M.,  {Nakajima} M.,  {Tashiro}
  M.~S.,    {Nakazawa} K.,  2009, \pasj, 61, 387

\bibitem[\protect\citeauthoryear{{Nugent}}{{Nugent}}{1998}]{Nugent1998}
{Nugent} R.~L.,  1998, \pasp, 110, 831

\bibitem[\protect\citeauthoryear{{{\"O}zel}}{{{\"O}zel}}{2013}]{Ozel2013}
{{\"O}zel} F.,  2013, Reports on Progress in Physics, 76, 016901

\bibitem[\protect\citeauthoryear{{Page}, {Geppert} \& {Weber}}{{Page}
  et~al.}{2006}]{Pageetal2006}
{Page} D.,  {Geppert} U.,    {Weber} F.,  2006, Nuclear Physics A, 777, 497

\bibitem[\protect\citeauthoryear{{Park}, {Hughes}, {Slane}, {Burrows}, {Lee} \&
  {Mori}}{{Park} et~al.}{2012}]{Parketal2012}
{Park} S.,  {Hughes} J.~P.,  {Slane} P.~O.,  {Burrows} D.~N.,  {Lee} J.-J.,
  {Mori} K.,  2012, \apj, 748, 117

\bibitem[\protect\citeauthoryear{{Park}, {Hughes}, {Slane}, {Mori} \&
  {Burrows}}{{Park} et~al.}{2010}]{Parketal2010}
{Park} S.,  {Hughes} J.~P.,  {Slane} P.~O.,  {Mori} K.,    {Burrows} D.~N.,
  2010, \apj, 710, 948

\bibitem[\protect\citeauthoryear{{Pires}, {Motch} \& {Janot-Pacheco}}{{Pires}
  et~al.}{2009}]{Piresetal2009}
{Pires} A.~M.,  {Motch} C.,    {Janot-Pacheco} E.,  2009, \aap, 504, 185

\bibitem[\protect\citeauthoryear{{Pons}, {Miralles} \& {Geppert}}{{Pons}
  et~al.}{2009}]{Ponsetal2009}
{Pons} J.~A.,  {Miralles} J.~A.,    {Geppert} U.,  2009, \aap, 496, 207

\bibitem[\protect\citeauthoryear{{Posselt}, {Popov}, {Haberl}, {Tr{\"u}mper},
  {Turolla} \& {Neuh{\"a}user}}{{Posselt} et~al.}{2007}]{Posseltetal2007}
{Posselt} B.,  {Popov} S.~B.,  {Haberl} F.,  {Tr{\"u}mper} J.,  {Turolla} R.,
   {Neuh{\"a}user} R.,  2007, \apss, 308, 171

\bibitem[\protect\citeauthoryear{{Posselt}, {Popov}, {Haberl}, {Tr{\"u}mper},
  {Turolla}, {Neuh{\"a}user} \& {Boldin}}{{Posselt}
  et~al.}{2010}]{Posseltetal2010}
{Posselt} B.,  {Popov} S.~B.,  {Haberl} F.,  {Tr{\"u}mper} J.,  {Turolla} R.,
  {Neuh{\"a}user} R.,    {Boldin} P.~A.,  2010, \aap, 512, C2

\bibitem[\protect\citeauthoryear{{Rutledge}, {Fox} \& {Shevchuk}}{{Rutledge}
  et~al.}{2008}]{Rutledgeetal2008}
{Rutledge} R.~E.,  {Fox} D.~B.,    {Shevchuk} A.~H.,  2008, \apj, 672, 1137

\bibitem[\protect\citeauthoryear{{Sasaki}, {Plucinsky}, {Gaetz} \&
  {Bocchino}}{{Sasaki} et~al.}{2013}]{Saakietal2013}
{Sasaki} M.,  {Plucinsky} P.~P.,  {Gaetz} T.~J.,    {Bocchino} F.,  2013, \aap,
  552, A45

\bibitem[\protect\citeauthoryear{{Sato}, {Bamba}, {Nakamura} \&
  {Ishida}}{{Sato} et~al.}{2010}]{Satoetal2010}
{Sato} T.,  {Bamba} A.,  {Nakamura} R.,    {Ishida} M.,  2010, \pasj, 62, L33

\bibitem[\protect\citeauthoryear{{Sawada} \& {Koyama}}{{Sawada} \&
  {Koyama}}{2012}]{SawadaKoyama2012}
{Sawada} M.,  {Koyama} K.,  2012, \pasj, 64, 81

\bibitem[\protect\citeauthoryear{{Schaab}, {Weber}, {Weigel} \&
  {Glendenning}}{{Schaab} et~al.}{1996}]{Schaabetal1996}
{Schaab} C.,  {Weber} F.,  {Weigel} M.~K.,    {Glendenning} N.~K.,  1996,
  Nuclear Physics A, 605, 531

\bibitem[\protect\citeauthoryear{{Schmidt}}{{Schmidt}}{1968}]{Schmidt1968}
{Schmidt} M.,  1968, \apj, 151, 393

\bibitem[\protect\citeauthoryear{{Schwope}, {Hasinger}, {Schwarz}, {Haberl} \&
  {Schmidt}}{{Schwope} et~al.}{1999}]{Schwopeetal1999}
{Schwope} A.~D.,  {Hasinger} G.,  {Schwarz} R.,  {Haberl} F.,    {Schmidt} M.,
  1999, \aap, 341, L51

\bibitem[\protect\citeauthoryear{{Seward} \& {Harnden} Jr.}{{Seward} \&
  {Harnden}}{1982}]{SewardHarnden1982}
{Seward} F.~D.,  {Harnden} Jr. F.~R.,  1982, \apjl, 256, L45

\bibitem[\protect\citeauthoryear{{Seward}, {Harnden} Jr. \& {Helfand}}{{Seward}
  et~al.}{1984}]{Sewardetal1984}
{Seward} F.~D.,  {Harnden} Jr. F.~R.,    {Helfand} D.~J.,  1984, \apjl, 287,
  L19

\bibitem[\protect\citeauthoryear{{Seward}, {Harnden} Jr., {Murdin} \&
  {Clark}}{{Seward} et~al.}{1983}]{Sewardetal1983}
{Seward} F.~D.,  {Harnden} Jr. F.~R.,  {Murdin} P.,    {Clark} D.~H.,  1983,
  \apj, 267, 698

\bibitem[\protect\citeauthoryear{{Shapiro} \& {Teukolsky}}{{Shapiro} \&
  {Teukolsky}}{1983}]{ShapiroTeukolsky1983}
{Shapiro} S.~L.,  {Teukolsky} S.~A.,  1983, {Black holes, white dwarfs, and
  neutron stars: The physics of compact objects}

\bibitem[\protect\citeauthoryear{{Smartt}}{{Smartt}}{2009}]{Smartt2009}
{Smartt} S.~J.,  2009, \araa, 47, 63

\bibitem[\protect\citeauthoryear{{Sun}, {Seward}, {Smith} \& {Slane}}{{Sun}
  et~al.}{2004}]{Sunetal2004}
{Sun} M.,  {Seward} F.~D.,  {Smith} R.~K.,    {Slane} P.~O.,  2004, \apj, 605,
  742

\bibitem[\protect\citeauthoryear{{Tendulkar}, {Cameron} \&
  {Kulkarni}}{{Tendulkar} et~al.}{2012}]{Tendulkaretal2012}
{Tendulkar} S.~P.,  {Cameron} P.~B.,    {Kulkarni} S.~R.,  2012, \apj, 761, 76

\bibitem[\protect\citeauthoryear{{Tetzlaff}, {Eisenbeiss}, {Neuh{\"a}user} \&
  {Hohle}}{{Tetzlaff} et~al.}{2011}]{Tetzlaffetal2011}
{Tetzlaff} N.,  {Eisenbeiss} T.,  {Neuh{\"a}user} R.,    {Hohle} M.~M.,  2011,
  \mnras, 417, 617

\bibitem[\protect\citeauthoryear{{Tetzlaff}, {Neuh{\"a}user}, {Hohle} \&
  {Maciejewski}}{{Tetzlaff} et~al.}{2010}]{Tetzlaffetal2010}
{Tetzlaff} N.,  {Neuh{\"a}user} R.,  {Hohle} M.~M.,    {Maciejewski} G.,  2010,
  \mnras, 402, 2369

\bibitem[\protect\citeauthoryear{{Tetzlaff}, {Schmidt}, {Hohle} \&
  {Neuh{\"a}user}}{{Tetzlaff} et~al.}{2012}]{Tetzlaffetal2012}
{Tetzlaff} N.,  {Schmidt} J.~G.,  {Hohle} M.~M.,    {Neuh{\"a}user} R.,  2012,
  \pasa, 29, 98

\bibitem[\protect\citeauthoryear{{Tiengo}, {Esposito}, {Mereghetti}, {Israel},
  {Stella}, {Turolla}, {Zane}, {Rea}, {G{\"o}tz} \& {Feroci}}{{Tiengo}
  et~al.}{2009}]{Tiengoetal2009}
{Tiengo} A.,  {Esposito} P.,  {Mereghetti} S.,  {Israel} G.~L.,  {Stella} L.,
  {Turolla} R.,  {Zane} S.,  {Rea} N.,  {G{\"o}tz} D.,    {Feroci} M.,  2009,
  \mnras, 399, L74

\bibitem[\protect\citeauthoryear{{Torra}, {Fern{\'a}ndez} \&
  {Figueras}}{{Torra} et~al.}{2000}]{Torraetal2000}
{Torra} J.,  {Fern{\'a}ndez} D.,    {Figueras} F.,  2000, \aap, 359, 82

\bibitem[\protect\citeauthoryear{{Turner}, {Rutledge}, {Letcavage}, {Shevchuk}
  \& {Fox}}{{Turner} et~al.}{2010}]{Turneretal2010}
{Turner} M.~L.,  {Rutledge} R.~E.,  {Letcavage} R.,  {Shevchuk} A.~S.~H.,
  {Fox} D.~B.,  2010, \apj, 714, 1424

\bibitem[\protect\citeauthoryear{{van Kerkwijk} \& {Kaplan}}{{van Kerkwijk} \&
  {Kaplan}}{2008}]{KerkwijkKaplan2008}
{van Kerkwijk} M.~H.,  {Kaplan} D.~L.,  2008, \apjl, 673, L163

\bibitem[\protect\citeauthoryear{{Vasisht} \& {Gotthelf}}{{Vasisht} \&
  {Gotthelf}}{1997}]{VasishtGotthelf1997}
{Vasisht} G.,  {Gotthelf} E.~V.,  1997, \apjl, 486, L129

\bibitem[\protect\citeauthoryear{{Vasisht}, {Kulkarni}, {Anderson}, {Hamilton}
  \& {Kawai}}{{Vasisht} et~al.}{1997}]{Vasishtetal1997}
{Vasisht} G.,  {Kulkarni} S.~R.,  {Anderson} S.~B.,  {Hamilton} T.~T.,
  {Kawai} N.,  1997, \apjl, 476, L43

\bibitem[\protect\citeauthoryear{{Vink}}{{Vink}}{2004}]{Vink2004}
{Vink} J.,  2004, \apj, 604, 693

\bibitem[\protect\citeauthoryear{{Voges}, {Aschenbach}, {Boller},
  {Br{\"a}uninger}, {Briel}, {Burkert}, {Dennerl}, {Englhauser}, {Gruber},
  {Haberl} \& {Hartner}}{{Voges} et~al.}{1999}]{Vogesetal1999}
{Voges} W.,  {Aschenbach} B.,  {Boller} T.,  {Br{\"a}uninger} H.,  {Briel} U.,
  {Burkert} W.,  {Dennerl} K.,  {Englhauser} J.,  {Gruber} R.,  {Haberl} F.,
  {Hartner} G.,  1999, \aap, 349, 389

\bibitem[\protect\citeauthoryear{{Walter} \& {Lattimer}}{{Walter} \&
  {Lattimer}}{2002}]{WalterLattimer2002}
{Walter} F.~M.,  {Lattimer} J.~M.,  2002, \apjl, 576, L145

\bibitem[\protect\citeauthoryear{{Walter}, {Wolk} \& {Neuh{\"a}user}}{{Walter}
  et~al.}{1996}]{Walteretal1996}
{Walter} F.~M.,  {Wolk} S.~J.,    {Neuh{\"a}user} R.,  1996, \nat, 379, 233

\bibitem[\protect\citeauthoryear{{Weisskopf}, {Elsner}, {Darbro}, {Leahy},
  {Naranan}, {Harnden}, {Seward}, {Sutherland} \& {Grindlay}}{{Weisskopf}
  et~al.}{1983}]{Weisskopfetal1983}
{Weisskopf} M.~C.,  {Elsner} R.~F.,  {Darbro} W.,  {Leahy} D.,  {Naranan} S.,
  {Harnden} F.~R.,  {Seward} F.~D.,  {Sutherland} P.~G.,    {Grindlay} J.~E.,
  1983, \apj, 267, 711

\bibitem[\protect\citeauthoryear{{Winkler} \& {Kirshner}}{{Winkler} \&
  {Kirshner}}{1985}]{WinklerKirshner1985}
{Winkler} P.~F.,  {Kirshner} R.~P.,  1985, \apj, 299, 981

\bibitem[\protect\citeauthoryear{{Wolszczan}, {Cordes} \& {Dewey}}{{Wolszczan}
  et~al.}{1991}]{Wolszczanetal1991}
{Wolszczan} A.,  {Cordes} J.~M.,    {Dewey} R.~J.,  1991, \apjl, 372, L99

\bibitem[\protect\citeauthoryear{{Yakovlev}, {Kaminker}, {Gnedin} \&
  {Haensel}}{{Yakovlev} et~al.}{2001}]{Yakovlevetal2001}
{Yakovlev} D.~G.,  {Kaminker} A.~D.,  {Gnedin} O.~Y.,    {Haensel} P.,  2001,
  \physrep, 354, 1

\bibitem[\protect\citeauthoryear{{Yakovlev} \& {Pethick}}{{Yakovlev} \&
  {Pethick}}{2004}]{YakovlevPethick2004}
{Yakovlev} D.~G.,  {Pethick} C.~J.,  2004, \araa, 42, 169

\bibitem[\protect\citeauthoryear{{Yar-Uyaniker}, {Uyaniker} \&
  {Kothes}}{{Yar-Uyaniker} et~al.}{2004}]{Yaretal2004}
{Yar-Uyaniker} A.,  {Uyaniker} B.,    {Kothes} R.,  2004, \apj, 616, 247

\bibitem[\protect\citeauthoryear{{Yuan}, {Wang}, {Manchester} \& {Liu}}{{Yuan}
  et~al.}{2010}]{Yuanetal2010}
{Yuan} J.~P.,  {Wang} N.,  {Manchester} R.~N.,    {Liu} Z.~Y.,  2010, \mnras,
  404, 289

\bibitem[\protect\citeauthoryear{{Zampieri}, {Campana}, {Turolla},
  {Chieregato}, {Falomo}, {Fugazza}, {Moretti} \& {Treves}}{{Zampieri}
  et~al.}{2001}]{Zampierietal2001}
{Zampieri} L.,  {Campana} S.,  {Turolla} R.,  {Chieregato} M.,  {Falomo} R.,
  {Fugazza} D.,  {Moretti} A.,    {Treves} A.,  2001, \aap, 378, L5

\bibitem[\protect\citeauthoryear{{Zane}, {Cropper}, {Turolla}, {Zampieri},
  {Chieregato}, {Drake} \& {Treves}}{{Zane} et~al.}{2005}]{Zaneetal2005}
{Zane} S.,  {Cropper} M.,  {Turolla} R.,  {Zampieri} L.,  {Chieregato} M.,
  {Drake} J.~J.,    {Treves} A.,  2005, \apj, 627, 397

\bibitem[\protect\citeauthoryear{{Zane}, {Haberl}, {Israel}, {Pellizzoni},
  {Burgay}, {Mignani}, {Turolla}, {Possenti}, {Esposito}, {Champion},
  {Eatough}, {Barr} \& {Kramer}}{{Zane} et~al.}{2011}]{Zaneetal2011}
{Zane} S.,  {Haberl} F.,  {Israel} G.~L.,  {Pellizzoni} A.,  {Burgay} M.,
  {Mignani} R.~P.,  {Turolla} R.,  {Possenti} A.,  {Esposito} P.,  {Champion}
  D.,  {Eatough} R.~P.,  {Barr} E.,    {Kramer} M.,  2011, \mnras, 410, 2428

\end{thebibliography}

\end{document}